# Superconducting Transformer for Superconducting Cable Testing up to 45 kA

H. Yu, and J. Lu

*Abstract*—A facility capable of testing superconducting cables with current of tens of kA is essential for the development of large superconducting magnets. A superconducting transformer (SCT) is a suitable choice as a high DC current source for testing superconducting cables. In this work, we will present our experimental results of a SCT up to output current of 45 kA. The properties of this SCT is characterized at 4.2 K. Its behaviors during quenches at different current levels are studied.

*Index Terms*—Superconducting transformer, superconducting cable, superconducting magnet, quench.

## I. Introduction

LARGE superconducting magnets systems require superconducting cables that carries tens of kA of electrical current. In order to develop and characterize large superconducting cables, it is necessary to apply electrical current at or above its operating current. Superconducting cable testing using a conventional DC power supply [1]-[2] requires large current leads and is energy-inefficient and expensive. A superconducting transformer (SCT) which uses a small current at its primary and obtains a large current output at its secondary, is a very efficient way to provide large current to superconducting cable samples.

In the past decades, a number of SCTs are developed for superconducting cable testing. For example, a 50 kA SCT is used as a current source at the SULTAN test facility at the Centre de Recherches en Physique des Plasma (CRPP) in Switzerland [3] and another SCT, which was designed to 50 kA and tested up to 38 kA, is used at FRESCA, the CERN cable test facility in Switzerland [4], [5]. The Lawrence-Berkley National Laboratory (LBNL) developed a 50 kA SCT and tested up to 28 kA [6]. The LBNL SCT was transferred in year 2016 to the National High Magnetic Field Laboratory USA (NHMFL) for enhancing NHMFL's cable testing capability in high magnetic fields.

This paper will present our characterization of this SCT up to 45 kA. The behaviors of the output current during a primary current ramp and especially during a quench are studied. The accuracy of the output current measurement will be discussed.

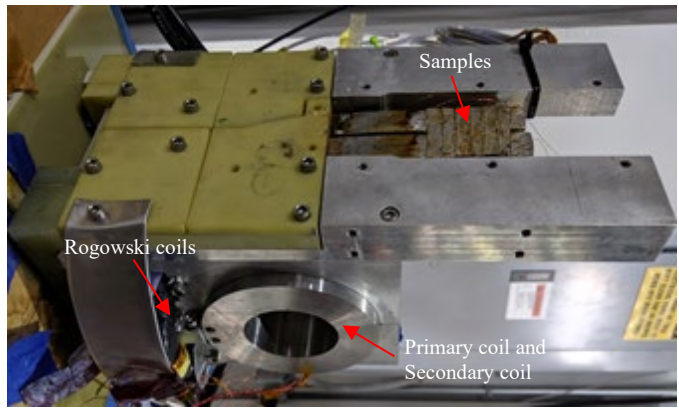

Fig. 1. A photograph of the SCT researched in this work.

## II. Experimental Methods

The SCT in this research consists of a primary coil made of NbTi wire, and a secondary coil made of NbTi Rutherford cable [6]. It equipped with two Rogowski coils for measuring output current, and two heaters near the output for quenching the SCT when needed. 50 turns of NbTi wire are wound on each Rogowski coil for its calibration by a reference current source. When the SCT was transferred to NHMFL, it was in a good condition, but the electronic control circuits as described in reference [6] was no longer available.

We first characterized the SCT's room temperature properties at the 100 Hz by using a HIOKI 3532-50 LCR meter. Table I listed the results and compared with the de-signed values. They are in reasonably good agreement.

The output current measurement is primarily by integrating the inductive voltage of a Rogowski coil that encircles an output current lead. This integration is done digitally by a computer within the data acquisition program written in LabVIEW. The voltages from Rogowski coil and various voltage taps are measured by a National Instruments DAQ voltage input module SCXI-1125.

The output current measurement by Rogowski voltage integration was calibrated 4.2 K by applying up to 80 A current to

This work is supported by the user collaboration grant program (UCGP) of the NHMFL which is supported by NSF through NSF-DMR-1157490 and 1644779, and the State of Florida.

H. Yu and J. Lu are with National High Magnetic Laboratory, Tallahassee-see, FL 32310, USA (corresponding author: J. Lu, junlu@magnet.fsu.edu).

Color versions of one or more of the figures in this paper are available online at http://ieeexplore.ieee.org.

Digital Object Identifier will be inserted here upon acceptance.

TABLE I
ROOM TEMPERATURE PROPERTIES MEASURED AND DESIGNED

| Room temperature properties | Measured (at 100 Hz) | Designed [6] |
|---|---|---|
| Current ratio | -- | 1000 |
| Primary inductance, $L_p$ (H) | 4.158 | 4.5 |
| Secondary inductance, $L_s$ (H) | 2.525E-6 | 2.27E-6 |
| Mutual inductance between primary and secondary, $M_{ps}$ (H) | 2.75E-3 | 2.8E-3 |
| Quench heater resistance (Ω) | 10.2 | -- |
| Rogowski inductance (H) | 8.257E-3 | 8.1E-3 |
| Mutual inductance between secondary and Rogowski (H) | 5.25E-6 | 5.3E-6 |
| Calibration inductance (H) | 2.253E-4 | -- |
| Mutual inductance between calibration and Rogowski (H) | 2.665E-4 | -- |
| Primary resistance, $R_p$ (Ω) | 636 | -- |

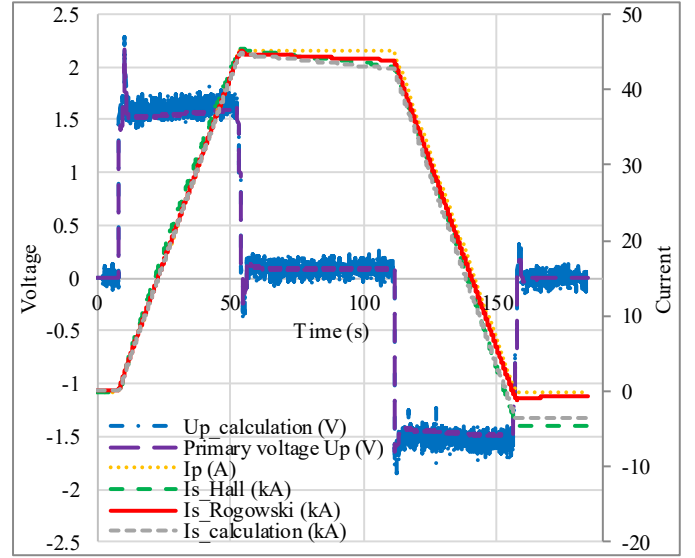

Fig. 3. A typical ramp and hold cycle at 45 kA. The unit of the primary current is 'A', while that of the secondary current is 'kA'.

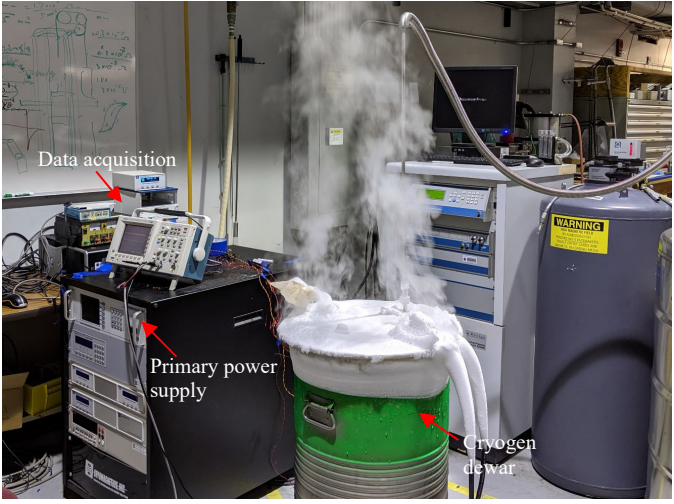

Fig. 2. A photo of the test system

the 50-turn calibration coil, which is equivalent to up to 4 kA output current. This was done by using a TDK Lambda GEN 8-180 DC power supply.

The output current measurement by inductive voltage integration is accurate if an initial current is known (presumably zero). For a SCT connected with a superconducting sample, however, the initial current can be significantly deviated from zero. This could result in a considerable error. Therefore for absolute output current measurement, albeit less accurate, we used two Hall sensors, one on the center of each output lead. The Hall sensors used were HZ-312C by Asahi Kasei Corporation and calibrated up to 1 tesla in a physical property measurement system made by Quantum Design. The output cur-rent was calculated by the measured magnetic field by Biot-Savart law.

In our experiments, SCT's output was shorted by 6 short pieces of superconducting cables soldered to the SCT output in parallel as shown in Fig. 1. Each of these pieces are Tevatron type NbTi Rutherford cable with its critical current of about 15 kA at 1 T [7]-[8]. Since at 50 kA, the magnetic field at the short cable is lower than 1 T, so 6 cables connected in parallel are adequate for 50 kA output with a healthy margin. A soldering procedure was developed, which uses Sn63Pb37 solder with help of a soldering fixture consists of cartridge heaters and a pair of thermocouple.

### III. RESULTS AND DISCUSSIONS

#### A. Maximum operating current

After cooling down the SCT to 4.2 K, the primary current was ramped up at 1 A/s. As the primary current linearly ramped up, as expected, the output current increased with a ratio of 1000:1 as shown in Fig. 3. The first spontaneous quench occurred at about 30 kA. Subsequent quenches (both induced and spontaneous in three cooldown campaigns) steadily increased quench current. In other words, the SCT clearly showed a training behavior similar to that in NbTi accelerator magnets. After about total of 20 quenches, its quench current reached 45.5 kA. Fig. 3 demonstrates a typical ramp-up, hold, and ramp-down cycle at 45 kA.

In order to better understand and predict the behavior of SCT, we carried out theoretical simulations. The currents and voltages in a transformer can be described by follow differential equations,

$$U_p = I_p R_p + L_p(dI_p / dt) - M_{ps}(dI_s / dt), \quad (1a)$$
$$0 = I_s R_s + L_s(dI_s / dt) - M_{ps}(dI_p / dt), \quad (1b)$$

where subscripted 'p' and 's' denote primary and secondary respectively. $L_p$, $L_s$, and $M_{ps}$ are respectively inductances of the primary, secondary, and mutual inductance between the primary and secondary. $L_p$, $L_s$, and $M_{ps}$ at room temperature are listed in Table I. The value of $R_s$ is essentially the resistance of the solder joint between SCT and the short cables, and a $R_s$ = 2.29 nΩ was obtained experimentally by fitting the $V_s$-$I_s$ data (not shown) with a straight line. So the secondary current $I_s$ can be calculated by Eq. (1b), which is plotted as $I_s\_calculation$ in Fig. 3.

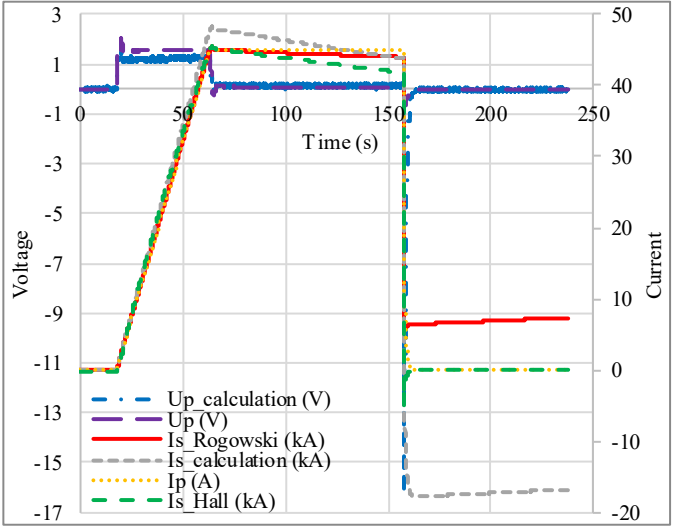

Fig. 4. An induced quench at 45 kA.

Fig.3 also shows that while the primary current is held at a constant level, the output current slowly drifts down. This is consistent with a predictable current decay due to the joint resistance between the output leads and the short cables. Our analysis shows that the decay time constant $\tau$ is about 3000 s consistent with Eq. (1b) where $\tau = L_s/R_s$.

It should be noted that as a consequence of the current decay, when the primary current was ramped back to zero, the output current became negative. This might have important implication in SCT application in critical current measurements, where the flux-flow resistance of the sample makes this decay much faster.

### B. Quench

The maximum energy stored in this SCT is estimated to be 45 kJ. This is equivalent to the energy for boiling off 2 liters of liquid helium. So a quench of the SCT will not post a serious safety concern. The potentially high inductive voltage during a quench, however, is a concern for the safety of the operator and the data acquisition instrumentations.

During experiments, quenches were induced by energizing the quench heaters attached to the secondary coil. In the cases of quench, the primary power supply detected the quench (The quench is detected when there is more than 4 A current drop for longer than 200 ms.) and automatically switch off its output. This induced quench event is depicted in Fig.4 where the current and voltages at both primary and secondary of the SCT are plotted.

It is noticed that the output current after quench cannot be measured correctly by Rogowski voltage integration. This seems to be because either the inductive voltage is out of the measurement range, or the data acquisition speed of 2 ms per data set used in our experiment is not fast enough to handle such a fast event. In such a case, the advantage of the Hall measurement is evident. Since the measured primary voltage $U_p$ during a quench is less than 2 V, the SCT operation is safe without additional protections.

TABLE II
MINIMAL QUENCH HEATER POWER NEEDED TO QUENCH THE SCT

| Nominal secondary current (kA) | Quench heater current (mA) | Minimal heater power (mW) |
|---|---|---|
| 45 | 120 | 144 |
| 20 | 220 | 488 |
| 10 | 280 | 784 |

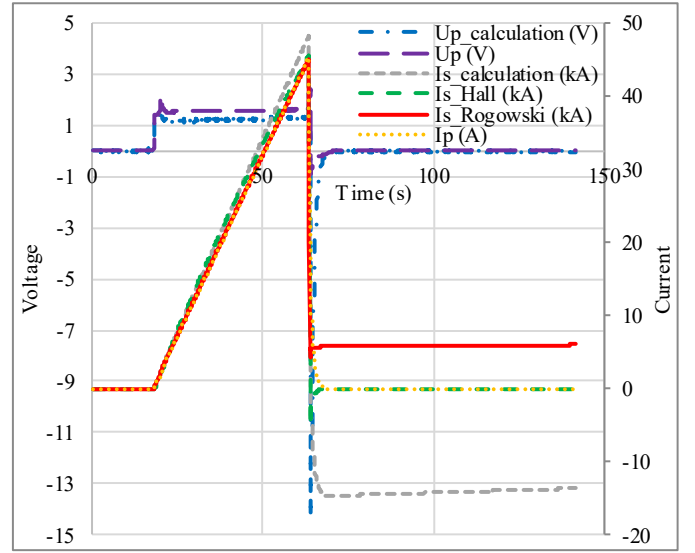

Fig. 5. Spontaneous quench at 45.5 kA.

The calculated primary voltage $U_p$ and the secondary current $I_s$ are also shown in Fig. 4. In this calculation, $R_s$ is assumed to increase rapidly at the moment of quench. The calculated $U_p$ spike of -16 V is not consistent with the measured value of -0.93V. Similarly, calculated $I_s$ of -17 kA does not agree with the measurements. The reason might be that the assumption that $I_p$ becomes zero at the moment of quench is not necessarily true. Measurements with much higher speed are needed to understand this discrepancy.

We also studied the minimum energy needed to induce a quench as a part of the SCT operation preparation. Table II shows the minimal quench heater power needed at three different secondary current levels. As expected, the power needed decreases with increasing operating current.

For the case of a spontaneous quench, Fig. 5 shows one that occurred at 45.5 kA. Since no superconducting-to-normal transition was observed on the short cables where there are voltage taps, the quench seemed to have been initiated in the SCT. Similar to above case of induced quench, $U_p$ spike was only -5 V, which was confirmed by a measurement by a digital oscilloscope with a time resolution of 100 μs.

Calculated $U_p$ and $I_s$ are also shown in Fig. 5. Comparison between calculation and measurement show similar discrepancy as in the case of induced quench.

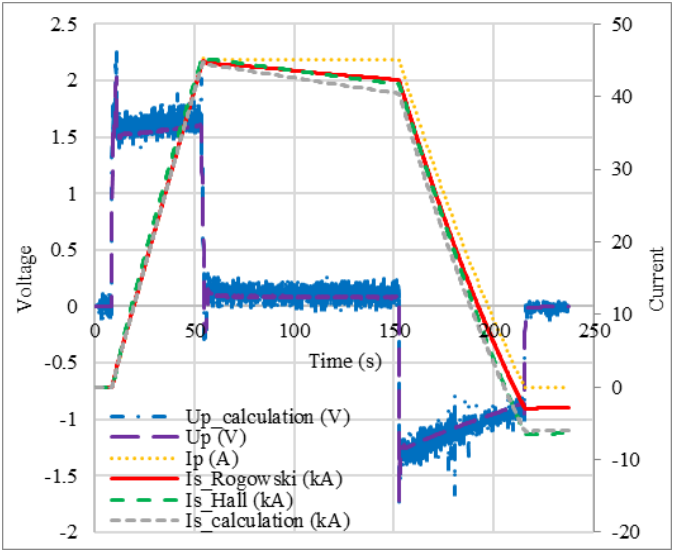

Fig. 6. Power supply shut down at about 150 s.

Finally the behavior of SCT was observed after a primary power supply shut down simulating the event of power outage, as it is a possible event in a real SCT operation. In this experiment, the SCT was ramped to 45 kA, then the main switch of the power supply was turned off. The results are displayed in Fig. 6. After the shut-down, the primary current did not became zero quickly. It decayed slowly at about 0.7 A/s. As a result, the primary inductive voltage is moderate. This behavior is likely due to the fact that primary power supply has low internal resistance after the shut-down. So the stored energy in SCT can be slowly discharged. The result of this experiment proved that a power-outage does not create a safety hazard to the SCT operation.

*C. Uncertainty in output current measurement*

It can be seen from Figs. 3-5 that $I_s$ measured by integration of Rogowski signal and by Hall sensors are in reasonably good agreement with each other except for those after a quench which we have discussed earlier. But small differences between the two can be distinguished. The main source of $I_s$ error comes from the Hall sensor measurements. This error can be broken down to the error in Hall sensor calibration, and the error in sensor position at 4.2 K which is used to calculated current by the measured magnetic field. In comparison, Rogowski signal integration measurement is not sensitive to geometrical uncertainties. However, in spite of its relatively larger measurement error, Hall measurement is crucial to obtain absolute output current when the initial current state is unknown, or when a fast event like a quench that makes Rogowski measurement unreliable, as demonstrated in Fig.4 and 5 after the quenches.

## IV. CONCLUSION

The SCT is characterized at 4.2 K at the NHMFL. It was successfully ramped and held up to 45 kA after a number of training quenches. The behavior of induced and spontaneous quenches were studied. Neither induced nor spontaneous quenches generated voltage higher than 5 V in the entire system. The SCT is safe to operate. Integration of Rogowski coil and Hall sensors are complimentary methods of measuring output current.

## ACKNOWLEDGMENT

We thank Dr. Hubertus Weijers for helpful discussions and Mr. Eric Stiers for helps on electronic control.

## REFERENCES

[1] D. C. van der Laan, P. D. Noyes, G. E. Miller, H. W. Weijers, and G. P. Willering, "Characterization of a high-temperature superconducting conductor on round core cables in magnetic fields up to 20 T," *Supercond. Sci. Technol.*, vol 26, Feb 2013, Art. no. 045005.
[2] M. Takayasu, L. Chiesa, P. D. Noyes, and J. V. Minervini, "Investigation of HTS twisted stacked-tape cable (TSTC) conductor for high-field, high-current fusion magnets," *IEEE Trans. Appl. Supercond.*, vol. 27, no. 4, Jun 2017, Art. no. 6900105.
[3] G. Pasztor, E. Aebli, B. Jakob, P. Ming, E. Siegrist, and P. Weymuth, "Design, construction and testing of a 50 kA superconducting transformer," *11th Int. Conf. Magn. Technol. (MT-11)*, 1990, pp. 467-472.
[4] A. P. Verweij, J. Genest, A. Knezovic, D. F. Leroy, J.-P. Marzolf, and L. R. Oberli, "1.9 K test facility for the reception of the superconducting cables for the LHC," *IEEE Trans. Appl. Supercond.*, vol. 9, no. 2, Jun 1999, pp. 153-156.
[5] A. P. Verweij, C-H Denarie, S. Geminian, and O. Vincent-Viry, "Superconducting transformer and regulation circuit for the CERN cable test facility," *J. Phys. Conf. Ser.*, vol 43, 2006, pp. 833–836.
[6] A. Godeke, D. R. Dietderich, J. M. Joseph, J. Lizarazo, S. O. Prestemon, G. Miller, and H. W. Weijers, "A superconducting transformer system for high current cable testing," *Rev. Sci. Instrum.*, vol 81, Mar 2010, Art. no. 035107.
[7] A. D. McInturff, R. A. Lundy, R. Remsbottom, and M. Wake, "I c (H, T) measurements for multi-kiloampere superconducting magnet conductor," *IEEE Trans. Magn.*, vol 21, no. 2, Mar 1985, pp. 975-978.
[8] L. Bottura, "A practical fit for the critical surface of NbTi," *IEEE Trans. Appl. Supercond.*, vol 10, no. 1, Mar 2000, pp. 1054-1057.